# Phononic transport in 1T'-MoTe$_2$: anisotropic structure with an isotropic lattice thermal conductivity

Xiangyue Cui [a], Xuefei Yan [b], Bowen Wang [a], Yongqing Cai [a*]

[a] Joint Key Laboratory of the Ministry of Education, Institute of Applied Physics and Materials Engineering, University of Macau, Taipa, Macau, China.
[b] School of Microelectronics Science and Technology, Sun Yat-sen University, Zhuhai, People's Republic of China.

* Corresponding author.
*E-mail address*: yongqingcai@um.edu.mo (Yongqing Cai)

**Abstract**

Molybdenum ditelluride (MoTe$_2$) is an unique transition metal dichalcogenide owing to its energetically comparable 1H and 1T' phases. This implies a high chance of coexistence of 1H/1T' heterostructures which poses great complexity in the measurement of the intrinsic lattice thermal conductivities ($\kappa_L$). In this work, via first-principles calculations, we examine the lattice-wave propagation and thermal conduction in this highly structurally anisotropic 1T' MoTe$_2$. Our calculation shows that the 1T' phase has a sound velocity of 2.13 km/s (longitudinal acoustic wave), much lower than that of the 1H phase (4.05 km/s), indicating a staggered transmission of lattice waves across the boundary from 1H to 1T' phase. Interestingly, the highly anisotropic 1T' MoTe$_2$ shows nearly isotropic and limited $\kappa_L$ of 13.02 W/mK, owing to a large Grüneisen parameter of acoustic flexural mode, heavy masses of Mo and Te elements and a low phonon group velocity. Accumulative $\kappa_L$ as a function of mean free path (MFP) indicates phonons with MFP less than ~300 nm contribute 80% of $\kappa_L$ and an inflection point at ~600 nm. Our results will be critical for understanding of the size dependent $\kappa_L$ of nanostructured 1T' MoTe$_2$.

# 1. Introduction

Since graphene - a carbon monolayer with a honeycomb two-dimensional (2D) lattice was first exfoliated from graphite in 2004 [1], extensive efforts have been devoted into exploring other 2D materials beyond graphene [2-5]. Transition metal dichalcogenides (TMDs) emerge recently as an unique crystalline system because of their tunable bandgap offers, strong electron-orbital-lattice coupling, and rich polymorph structures [6]. Molybdenum ditelluride (MoTe$_2$) as a typical versatile TMDs material exists three stable crystal structures: 1H, 1T' and T$_d$ phase at ambient condition without the need of chemical additions as required for other TMDs such as MoS$_2$. In particular, the coexistence of semiconducting 1H phase and metallic 1T' phase allows the fabrication of coplanar metal-semiconducting heterostructures or quantum dot with ultra-low contact conductance of 0.003 $G_0$ ($G_0 = 2e^2/h$) [7] for a wide range of applications. The class of TMDs materials is also an enticing platform for the fields of optoelectronic devices [8-11], field effect transistors [12-14], spintronic and quantum computational [15,16], low-resistance contacts and phase-change memory [17]. Recently, stability of monolayer 1T'-$MoTe_2$ under environmental conditions has been dramatically improved via passivating hexagonal boron nitride (hBN) above the newly grown 1T' phase, and its lifetime has been prolonged to more than a month [18], which sets a solid background for storage and applications of 1T'-$MoTe_2$.

Strain engineering was proven to be an effective way to modulate the physical properties of various 2D materials via inducing changes in band structure and atomic bond configurations [19], so as to influence the electronic and optical properties [20-24] as well as lattice vibration [23,25]. Amongst various TMDs, the $MoTe_2$ shows the smallest difference (~ 35 meV per unit cell) in the free energy between the semiconducting 1H phase and metallic 1T' phase [26,27]. The 1H-$MoTe_2$ was found to transform into 1T' phase under a small uniaxial strain of 0.2% generated via an atomic force microscope tip [28]. A recent study analyzed the energy landscape and predicted that strain and doping engineering can further alter the 1H-1T' phase diagram of the $MoTe_2$ [7]. Yang *et al*. [25] systematically studied the vibrational responses of single-limit of 1H-$MoTe_2$ under equibiaxial strain, and found that acoustic branch softens at the K point under 11.27% compressive strain originating from the Fermi surface nesting. Besides that, effects of strain on lattice thermal conductivity [29,30], Raman spectrum [31,32] and magnetic properties [33] of materials have also been reported. In spite of extensive researches on the structural and electronic properties, however, evolutions of phonon thermal transport in 1T'-$MoTe_2$ with lattice deformation remains unclarified.

Phonons as a type of quasiparticle are in form of the lattice vibration, and affect the stability, thermal conduction and lifetime of carriers of the hosting lattice. Dilation of lattice would alter the bonding strength and thus modulating the phonons and renormalizing the electronic and transport properties. In this paper, we explore the lattice dynamics under biaxial strains and thermal transport of monolayer 1T' phase of MoTe$_2$. Through evaluating the anharmonic properties, for example, the phonon group velocity ($v$), Grüneisen parameter ($\gamma$), phonon relaxation time ($\tau$) and lattice thermal conductivity ($\kappa_L$), the anharmonic properties and relaxation time of lattice waves for the 1T' MoTe$_2$ are illustrated. Surprisingly, unlike its isostructural 1T'-WTe$_2$ [34], the thermal conduction of monolayer 1T'- MoTe$_2$ shows a weaker anisotropy. The γ of ZA mode in the low frequency appear strongly negative, which is dramatically different from that of MoS$_2$ [35]. We also reveal the cumulative $\kappa_L$ via phononic mean free path which allows us to quantitatively estimate the size-dependent behavior of the $\kappa_L$.

## 2. Methods

Our calculations were performed by using the Quantum ESPRESSO package [36] based on first-principles calculations with density functional perturbation theory (DFPT) [37]. We used norm-conserving pseudopotentials within local density approximation (LDA) functional from Hartwigesen-Goedecker-Hutter PP table. The thickness of the vacuum region is about 15 Å to avoid spurious interactions and the setting of plane-wave cutoff energy is 100 Ry. The Brillouin zone (BZ) integration was adopted a Gaussian smearing of 0.01 Ry with a Monkhorst-Pack $k$-mesh of 10 × 18 × 1 (16 × 16 × 1) for 1T'-MoTe$_2$ (1H-MoTe$_2$). The criteria for the convergence of the total energy and Hellmann-Feynman force are $10^{-8}$ ($10^{-6}$) Ry and $10^{-7}$ ($10^{-5}$) Ry/bohr, respectively.

For the phonon calculations, a 5 × 9 × 1 (8 × 8 × 1) $q$-grid was used to obtain the interatomic force constants (IFC) of 1T'-MoTe$_2$ (1H-MoTe$_2$) from calculating dynamical matrix at each $q$ point through Fourier transformation. The equibiaxial strain used in this paper is expressed as $\varepsilon = [(a - a_0) / a_0] \times 100\%$, where $a$ and $a_0$ represent the strained and pristine lattice constant, respectively. Positive values of $\varepsilon$ represent tensile strains and negative ones correspond to compressive strains. As for the anharmonicity of phonons, phonon-phonon scattering rate and lattice thermal conductivity, were derived by iteratively solving Boltzmann transport equation for phonons within ShengBTE package [38]. The $\kappa_L$ can be calculated as follows

$$\kappa_L = \frac{1}{V} \sum_\lambda C_\lambda v_{\lambda\alpha} v_{\lambda\beta} \tau_\lambda \tag{1}$$

where $V$ and $\lambda$ represent crystal volume and phonon modes with different wave vectors and branches, respectively. The $C_\lambda$ is specific heat for each mode, indicating the capacity of heat can be stored in solids by generating acoustic and optical phonons [39]; $v_{\lambda\alpha}$ and $v_{\lambda\beta}$ are the group velocity along $\alpha$ and $\beta$ directions, respectively; $\tau_\lambda$ is relaxation time with different phonon mode $\lambda$, which is associated with anharmonic properties of phonons. Here, we calculate three phonon-phonon interaction related anharmonic force constant (3IFC) with a 2 × 4 × 1 supercell and the 12 × 24 × 1 q-points over the Brillouin zone.

## 3. Results and discussion

The monolayer 1T' (1H) phase of MoTe$_2$ is modeled with a distorted monoclinic (honeycomb) crystal structure, and the relaxed lattice constants are $a$ = 6.29 Å, $b$ = 3.39 Å ($a$ = $b$ = 3.48 Å), smaller than the previous results [40] as the LDA method leads to underestimated values. The unit cell of monolayer 1H and 1T' phase has three and six atoms, respectively, and there are a total of 9 and 18 phonon modes accordingly. In the following, since the phononic properties of 1T' phase are less explored than its 1H counterpart, we will mainly focus on the phonon properties of 1T' phase and a proper comparison with 1H phase is added if necessary. The 1T'-MoTe$_2$ structure is intrinsically anisotropic with the space group $P2_1/m$ and the point group $C_{2h}$ (2/m), which belongs to centrosymmetric structure and possesses 4 irreducible representations. Based on factor group analysis, its phonon irreducible representation at Γ point can be expressed as following:

$$\Gamma = [6A_g + 3B_g] + [2A_u + 4B_u] + [A_u + 2B_u] \tag{2}$$

where the first term represents Raman active mode, the second term represents infrared-active mode, and the third one represents acoustic mode, respectively. The symbols of *A* and *B* indicate single degenerate state symmetrical or antisymmetric to the principal axis, respectively, and subscripts *u* and *g* represent the mode is antisymmetric and symmetric to the operation of inversion symmetry.

For comparison, we firstly calculated the phonon dispersion of monolayer 1H phase of MoTe$_2$ as shown in Fig. 1(a), there are 3 acoustic branches: out-of-plane acoustic (ZA), transverse acoustic (TA), longitudinal acoustic (LA) and 6 optical branches with none imaginary frequency of each branch indicating the dynamical stability. Among them, the ZA mode is parabolic near the Γ point, which is a typical feature of the phonon spectrum of two-dimensional materials [41,42]. Phonon group velocity *v*, as one of the important physical quantities to describe harmonic properties, is determined from the slope of the phonon dispersion, and calculated as $v = d\omega/dq$, where *ω* and *q* represent frequency and wave vector of phonons, respectively. In Fig. 1(b), we plotted the variation of *v* with *ω* along Γ to M direction (the results of group velocity along Γ to K are shown in the Fig. S1). The highest phonon group velocity of TA (*v*-TA) and LA (*v*-LA) mode is about 2.52 and 4.05 km/s which is located at the Γ point, in line with the result of Shafique *et al.* [30], much lower than that of graphene [43]. From the figure, we also can see clearly that the *v*-TA and *v*-LA decreases with the *ω*, while that of value of ZA branch initially increases, then reaches a maximum value of ~ 2.70 km/s and drops sharply to the minimum at 106.6 cm$^{-1}$.

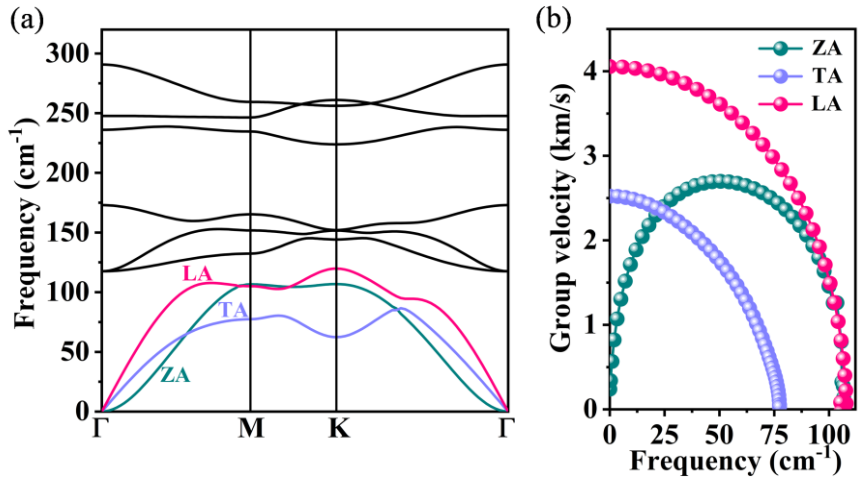

**Fig. 1.** (a) Phonon dispersion curves of monolayer 1H-MoTe$_2$, and (b) phonon group velocity of pristine 1H phase for ZA, TA and LA mode along Γ to M direction.

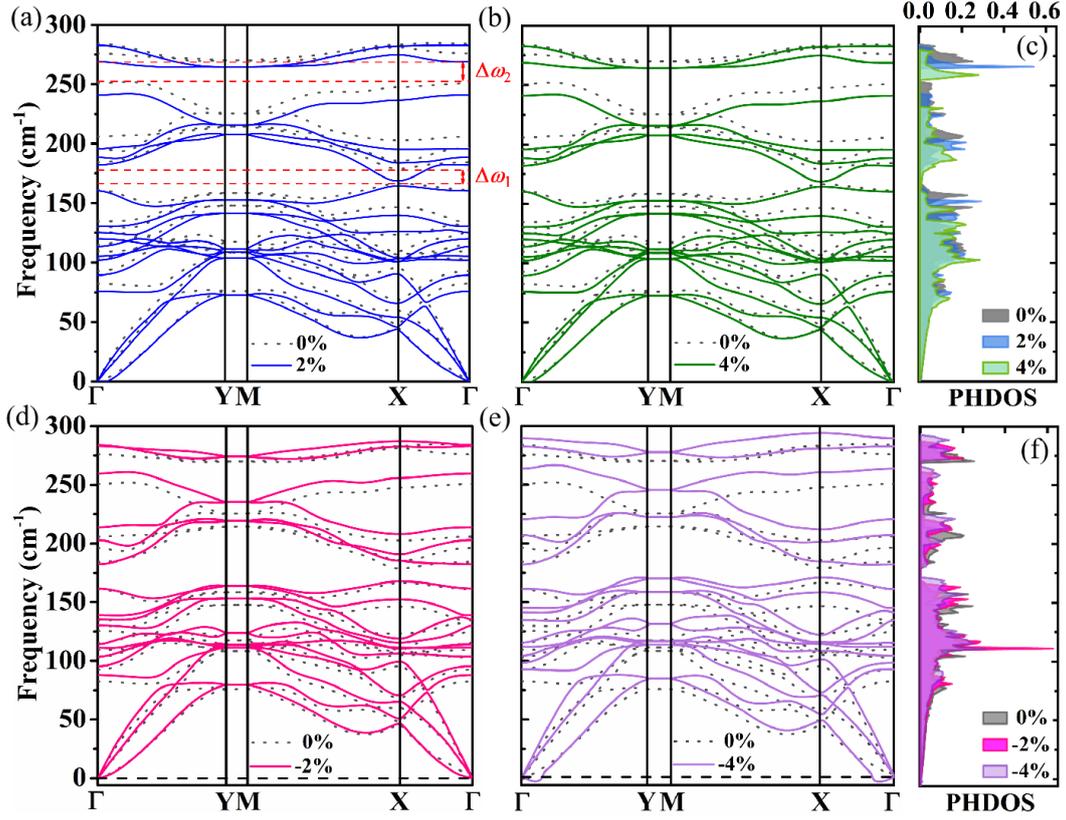

**Fig. 2.** Phonon dispersion curves of monolayer 1T'-MoTe$_2$ along high-symmetry directions and phonon density of states (PHDOS) under different biaxial strain. The tensile strains of (a) 2%, (b) 4% and corresponding the PHDOS. (c) The compressive strain of (d) -2%, (e) -4% and the PHDOS (f). The black dash curves represent phonon spectrum of pristine 1T'-MoTe$_2$, which is compared with the counterpart of strained ones.

The information of lattice dynamics of monolayer 1T' MoTe$_2$ is crucial to precisely estimate its phonon transport properties, and set as the foundation for discussing the effects of moderate equibiaxial strain. As depicted in Fig. 2, the phonon dispersion of monolayer 1T'-MoTe$_2$ under zero strain shows superior dynamical stability with the absence of imaginary frequency, and the frequencies for all optical phonon modes at Γ point are compiled in Table 1. The highest phonon frequency is only ~284.0 cm$^{-1}$, much lower than that of MoS$_2$ [35], indicating a lower phonon group velocity in MoTe$_2$.

As expected, there are minor imaginary modes developed for the ZA branch under compressive strain (-4%), reflecting the structural instability and tendency of rippling for such 2D systems. Nevertheless, as pointed out by Yang *et al.* [25], the issue could be partially suppressed by the interaction between the film and the substrate when put 2D material on a substrate in experiment. For the optical branches, the frequencies generally undergo red shift (blue shift) under tensile (compressive) strains in Fig. 2. Especially the high-frequency branches of $B_u^3$ (Raman active), $A_g^5$ (Raman active) and $A_g^6$ (Infrared active) change obviously, and their vibrational patterns are plotted in Fig. S2. The vibration amplitudes of Mo atoms of these modes are significantly larger than that of Te atoms which can be associated with the lighter mass of Mo atoms which are more easily activated.

There are phononic gaps formed between low-frequency bands (82.3 - 166.5 cm$^{-1}$), medium-frequency bands (178.9 - 252.0 cm$^{-1}$) and high-frequency band (269.5 - 284.5 cm$^{-1}$) estimated as $\Delta\omega_1$ = 12.4 cm$^{-1}$ and $\Delta\omega_2$ = 17.5 cm$^{-1}$, respectively (Fig. 2a). Upon applying tensile strains, the $\Delta\omega_1$ narrows significantly and the $\Delta\omega_2$ broadens with strain. In contrast, the $\Delta\omega_1$ initially increases slightly at -2% strain and then reduces again at -4% compressive strain, while the $\Delta\omega_2$ keeps decreasing under compressive strain. These modulated gaps largely depend on the considerable change of vibration frequencies between atoms under different strains, and ultimately dated back to the remarkable bonding variations as indicated by the continually changed separation in vertically aligned Te atoms ($d_1$ and $d_2$) shown in Fig. S3. Different from the 1H phase (Fig. 1(a)), no obvious phononic gap between the acoustic and optical branches is observed in 1T'-MoTe$_2$, and this continual distribution of frequencies increases the chance of phonon-phonon scattering in 1T'-MoTe$_2$ as the criteria of energy conservation is ensured by low-frequency acoustic modes.

**Table 1**

The phonon frequencies (cm$^{-1}$) and Grüneisen parameters ($\gamma$) of optical phonon modes at the Γ point for pristine 1T'-MoTe$_2$.

| symmetry | Exp. [18] | Cal. [40] | Cal. [44] | Cal. (this work) | $\gamma$ |
|---|---|---|---|---|---|
| $A_g^1$ | 80.41 | 80.4 | 82.5 | 82.4 | 1.81 |
| $B_g^1$ | 102 | 73.8 | 86.3 | 92.7 | 0.81 |
| $B_g^2$ | 163 | 101.0 | 98.3 | 103.0 | 0.19 |
| $A_u^1$ | | 105.8 | 104.2 | 109.5 | 0.69 |
| $A_g^2$ | 113.84 | 111.3 | 112.2 | 115.1 | 0.09 |
| $B_u^1$ | | 121.7 | 124.6 | 125.6 | 1.06 |
| $A_g^3$ | 128 | 125.7 | 127.6 | 130.0 | 0.99 |
| $B_u^2$ | | 130.9 | 135.2 | 134.7 | 0.79 |
| $A_g^4$ | - | 161.1 | 160.6 | 161.2 | 0.04 |
| $A_u^2$ | | 179.7 | 182.0 | 184.5 | -0.02 |
| $B_g^3$ | 188 | 162.5 | 195.8 | 196.8 | 0.95 |
| $B_u^3$ | | 198.6 | 207.4 | 205.9 | 1.10 |
| $A_g^5$ | 252 | 254.6 | 254.2 | 250.8 | 0.89 |
| $A_g^6$ | 269 | 269.2 | 276.8 | 276.4 | 0.60 |
| $B_u^4$ | | 278.6 | 285.0 | 284.1 | 0.07 |

Based on the phonon dispersions, we calculated harmonic - group velocities, one of the key quantities for deriving phonon thermal transport. The values of $v$ for acoustic phonons in intrinsic 1T' phase along Γ-Y direction are 1.47 (2.13) km/s, and 1.75 (2.30) km/s in Γ-X direction for TA (LA) branch, respectively, much lower than those of the 1H phase (the details are shown in Table S1). Furthermore, the strain effect on relationship between $v$ and frequency ($\omega$) of 1T' phase along Γ-Y direction are plotted in Fig. 3 (the counterpart along Γ-X direction are shown in Fig. S4). The frequency position of peaks of velocity for the three acoustic phonon branches generally moves downward under tensile strain, but upward under compressive strain. As for in-plane acoustic TA and LA modes, although the group velocities almost exhibit downward trend continually with equibixial strains, the velocities increase under compressive strains due to the bonding weakening and strengthening. The detailed strain-dependent phonon group velocities of 1T' phase at Γ point corresponding to the long-wavelength acoustic waves are compiled in Table 2, which verify the conjecture of low phonon group velocity.

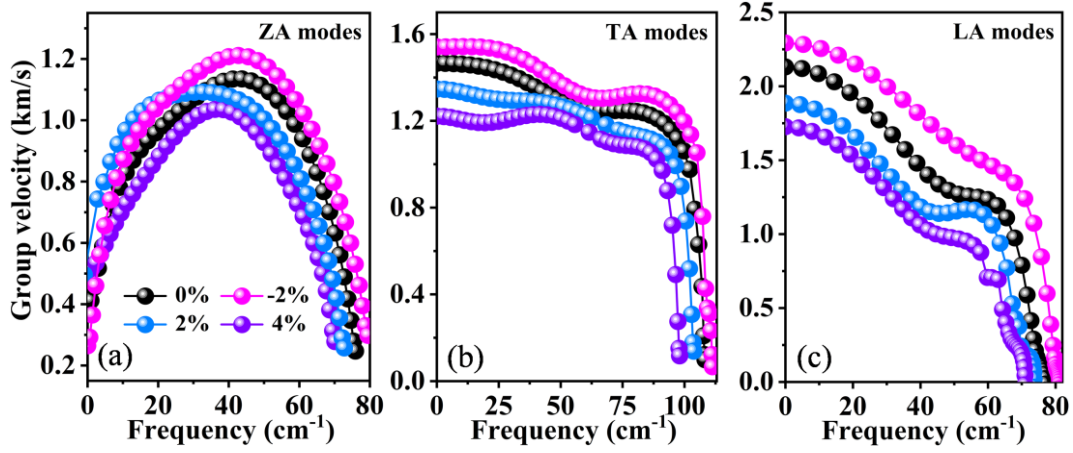

**Fig. 3.** The strain-dependent phonon group velocities for acoustic modes in monolayer 1T'-MoTe$_2$ along Γ-Y direction.

**Table 2**
The group velocities at Γ point for the TA ($v$-TA) and LA ($v$-LA) branches along Γ-Y direction of the pristine and strained monolayer 1T'-MoTe$_2$, where the positive numbers represent tensile strain, and the reverse is the compressive strain.

| strain | $v$-TA (km/s) | $v$-LA (km/s) |
| --- | --- | --- |
| 4% | 1.22 | 1.72 |
| 3% | 1.26 | 1.70 |
| 2% | 1.33 | 1.80 |
| 1% | 1.38 | 1.99 |
| 0% | 1.47 | 2.13 |
| -1% | 1.49 | 2.17 |
| -2% | 1.54 | 2.28 |

To deeply understand the phonon properties of monolayer 1T'-MoTe$_2$, we further calculated the $\gamma$ that normally represents the variation of phonon frequencies with material volume (here is the area for 2D 1T'-MoTe$_2$), and directly reflects the intensity of phonon anharmonicity. The $\gamma$ can be obtained by dilating the lattice with ± 0.5% strain of biaxial strains as defined by [35]

$$\gamma_q = -\frac{a_0}{2\omega_0}\frac{d\omega(q)}{da} \tag{3}$$

where $a_0$ is the equilibrium lattice constant of 6.290 Å, $\omega_0$ is the frequency of specifies phonon modes (*eg*: ZA, TA, LA). The computational results of $\gamma$ for acoustic modes at $\Gamma$ point are shown in Fig. 4. As we all know, most of materials normally expand with heating up and contract with cooling down, corresponding to a positive $\gamma$. Interestingly, the ZA mode of 1T'-MoTe$_2$ exhibits anomalous negative $\gamma$ ($\gamma$ = -2.96) along $\Gamma$-Y direction toward the zone center. The phenomenon is rare and abnormal, suggesting its area will shrink on heating which is quite different from some other 2D materials, i.e., graphene [45], T$_d$-MoTe$_2$ [42], and InX (X = S, Se, Te) [41]. We may elucidate that 1T'-MoTe$_2$ monolayer has a better ductility because of thermal shock resistance of a material is inversely proportional to its coefficient of thermal expansion [46]. However, toward the $\Gamma$-X direction, the $\gamma$ is positive ($\gamma$ = 26.38).

Above asymmetric behavior of $\gamma$ in ZA branch is mitigated for the TA and LA as shown in Fig. 4 and the $\gamma$ is 1.15 and 1.86, respectively. Both values are much lower than other TMDs materials with a high lattice thermal conductivity [35]. This indicates that the acoustic modes of 1T'-MoTe$_2$ tend to have a weaker anharmonicity, since the $\kappa_L$ is proportional to anharmonic interaction matrix elements (closely related to the $\gamma$). The relatively weak anharmonicity also implies a prolonged phonon lifetime since the phonon-phonon scattering rates depend on anharmonicity of the material [41]. For the optical branches, most of the $\gamma$ are lower than that of acoustic branches as listed in Table 1, indicated that they have a weaker phonon anharmonicity. Notably, there are three optical modes showing significant positive values of $\gamma$, including the $A_g^1$ (82.4 cm$^{-1}$, $\gamma$ = 1.81), $B_u^1$ (125.6 cm$^{-1}$, $\gamma$ = 1.06), and $B_u^3$ (205.9 cm$^{-1}$, $\gamma$ = 1.10). It is desired to confirm the predicted behaviors of these modes by potential electromechanical measurement.

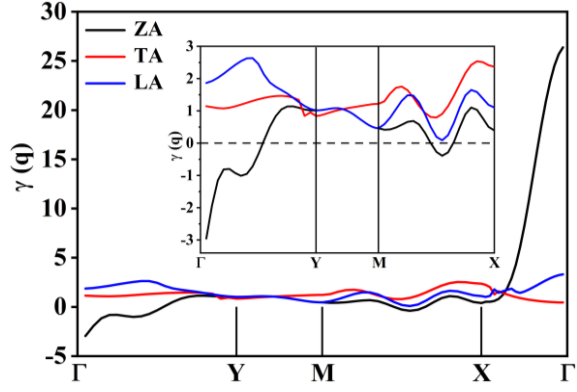

**Fig. 4.** The Grüneisen parameter γ of acoustic branch of monolayer 1T' phase of MoTe$_2$. The inset is corresponding close-up view partly.

Finally, we examine the key metrics for describing the thermal transport, that is the phonon relaxation time ($\tau$) and its reciprocal phonon scattering rates ($1/\tau$). Evaluation of $\tau$ needs to consider various scattering processes in real materials, including phonon-phonon scattering which is intrinsic and other scatterings, such as impurity scattering, boundary scattering, defect scattering which are more or less related to quality of sample. Herein, we only focus on the intrinsic phonon-phonon scattering at room temperature (300 K), which is closely associated with Umklapp process. The scattering rates and relaxation time of phonons are plotted in Fig. 5. Interestingly, the calculated scattering rates of 1T' MoTe$_2$ in the low frequency region are even higher than the well-known 2D materials with ultra-low lattice thermal conductivity [47]. The reason is twofold: firstly, the frequencies of lattice vibrations of the 1T' MoTe$_2$ is overall low (< 290 cm$^{-1}$), implying the lattice is very soft and the vibrations of the 2D lattice will be more easily scattered by the low-energy acoustic modes. Secondly, the acoustic and low-frequency optical bands have no gap and are strongly overlapped in phonon spectrum (Fig. 2), thus promoting the chances of multi-phonons scattering as more emission or absorption events will be allowed via low-energy phonons. Indeed, as shown in Fig. 5(a), there is a peak in the scattering rate of low-energy optical modes at around 100 cm$^{-1}$. Notably, the $\tau$ of three acoustic modes decreases sharply with phonon frequency in Fig. 5(b).

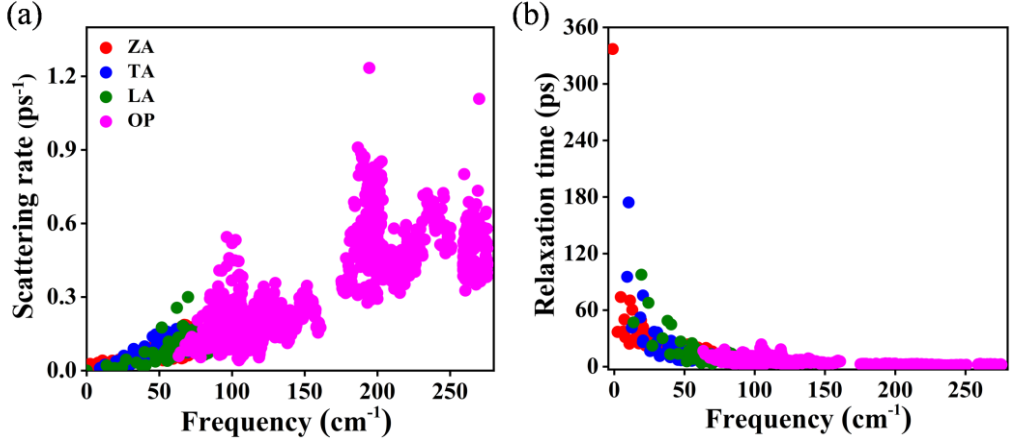

**Fig. 5.** The frequency-dependent (a) phonon scattering rates and (b) relaxation time for acoustic (ZA, TA, LA) and total optical (OP) modes of monolayer 1T'-MoTe$_2$.

The predicted intrinsic lattice thermal conductivity ($\kappa_L$) of 1T'-MoTe$_2$ monolayer as a function of temperature ($T$) is presented in Fig. 6(a). The $\kappa_L$ fits well with reciprocal of temperature ($1/T$), and the result elucidates that Umklapp process makes the main contribution to the three-phonon scattering process. At room temperature (300 K), the $\kappa_L$ is 13.25 and 13.02 W/mK along $x$ and $y$ directions, respectively. Specifically, the contributions of ZA, TA and LA modes to the $\kappa_L$ are 36%, 15%, 39% and 50%, 21%, 18% in the $x$ and $y$ directions, respectively. The acoustic phonons contributes greatly to the $\kappa_L$ because they characterize the motion of unit cell centroid, which vibrate in the same direction together with a higher group velocity compared with optical modes, thus facilitating thermal transport. As for the small difference of $\kappa_L$ between the two directions, it can be understood from the weak anisotropic phonon group velocity.

In fact, samples have a finite size in practical applications, where effect of the size on the $\kappa_L$ is significant due to additional boundary scattering, especially in nanoscale devices or at low temperature [34]. We next investigate phonon mean free path (MFP)-dependent cumulative lattice thermal conductivity as illustrated in Fig. 6(b). The cumulative $\kappa_L$ increases with phonon MFP and saturates at ~1232 nm (1.23 μm). Our results show that phonons with MFP less than ~300 nm contribute 80% of thermal conductivity. When the sample size decreases from the maximum MFP, the increase of phonon boundary scattering will lead to the decrease of $\kappa_L$ and weakening of the anisotropy. Assuming the typical sample size is ~1 μm, the $\kappa_L$ would be 13.16

W/mK and 12.79 W/mK in the *x* and *y* direction, respectively, which is based on the completely diffusive boundary condition in Fig. 6(b). Our calculations show that reduction of $\kappa_L$ at this size is limited because the saturating MFP of the system is relatively short.

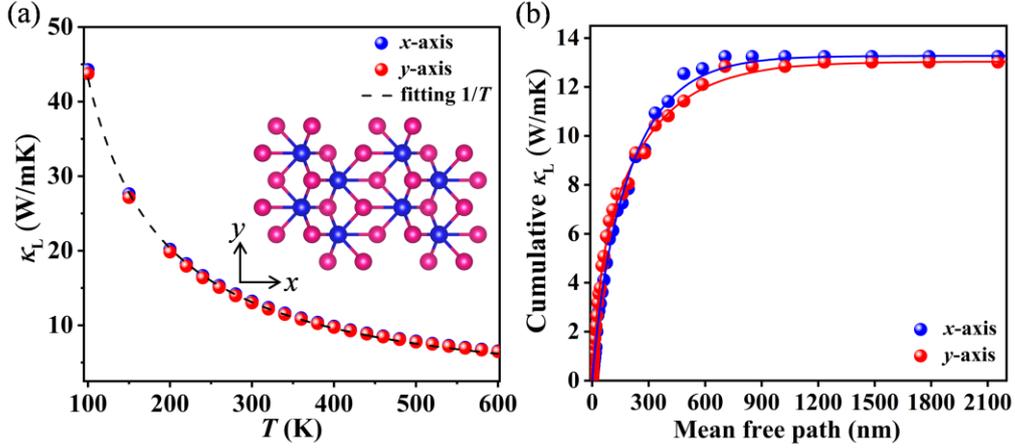

**Fig. 6.** (a) The temperature-dependent thermal conductivity, where dash line is the $1/T$ fitting results. An atomic model of 1T' phase in top view is shown in inset, the blue and pink spheres represent Mo and Te atoms, respectively. (b) Cumulative lattice thermal conductivity as a function of phonon mean free path for monolayer 1T'-MoTe$_2$ along *x* (blue symbols) and *y* (red symbols) directions at 300 K.

## 4. Conclusion

In summary, the anharmonic properties of the lattice vibrations in monolayer 1T'-MoTe$_2$ are explored by using the DFPT method. The sound velocities in the 1T' phase are much slower than its cousin 1H phase, implying a slower propagation of the acoustic waves. Our work suggests that those long-wavelength acoustic lattice vibrations have a much longer relaxation time up to 300 picosecond than optical lattice waves. The maximum frequency of the lattice waves is smaller than 290 cm$^{-1}$, indicating a very soft lattice of the 1T' phase which is likely to ripple. The rippling tendency is also implied by our calculated negative Grüneisen parameter for the ZA branch toward the long-wavelength limit. We also shows a weak anisotropic lattice thermal conductivity in the highly anisotropic 1T' phase. Our predicted cumulative lattice thermal conductivity indicates that the thermal conductivity becomes saturated at around 1.23 μm. This finding will be critical for the thermoelectric and nanoelectronic applications where the trend of size dependent thermal conductivity of 1T'-MoTe$_2$ nanostructures can be estimated.

## Data availability

The data that support the findings of this study are available upon reasonable request from the corresponding author.

## Conflict of interest

The authors declare no conflict of interest.

## CRediT authorship contribution statement

**Xiangyue Cui:** Investigation, Data curation, Formal analysis, Writing - original draft, Validation. **Xuefei Yan:** Formal analysis, Validation. **Bowen Wang:** Validation. **Yongqing Cai:** Supervision, Formal analysis, Funding acquisition, Writing - review & editing.

## Acknowledgments

This work was supported by the Natural Science Foundation of China (Grant 22022309), and the Natural Science Foundation of Guangdong Province, China (2021A1515010024), the University of Macau (SRG2019-00179-IAPME, MYRG2020-00075-IAPME), the Science and Technology Development Fund from Macau SAR (FDCT-0163/2019/A3). This work was performed at the High Performance Computing Cluster (HPCC), which is supported by the Information and Communication Technology Office (ICTO) of the University of Macau.